\documentclass[prl,aps,twocolumn,superscriptaddress]{revtex4-1}
\usepackage{latexsym}
\usepackage{graphicx}
\usepackage{multirow}
\usepackage{verbatim}
\usepackage{amsmath}

\newcommand{\be}{\begin{equation}}
\newcommand{\ee}{\end{equation}}
\newcommand{\ba}{\begin{eqnarray}}
\newcommand{\ea}{\end{eqnarray}}
\newcommand{\baa}{\begin{eqnarray*}}
\newcommand{\eaa}{\end{eqnarray*}}

\def\be{\begin{equation}}
\def\ee{\end{equation}}
\def\bea{\begin{eqnarray}}
\def\eea{\end{eqnarray}}

\def\C60{A$_x$C$_{60}$}

\def\HgCu3{HgCa$_2$Cu$_3$O$_{8+y}$}
\def\HgCu4{HgBa$_2$Ca$_3$Cu$_4$O$_{10+y}$}
\def\TlCu{Tl$_2$Ba$_2$CuO$_{6+\delta}$}
\def\TlCu3{Tl$_2$Ba$_2$Ca$_2$Cu$_3$O$_{10+y}$}
\def\TlCu4{Tl$_2$Ba$_2$Ca$_3$Cu$_4$O$_{12+y}$}

\def\BiCu3{Bi$_2$Sr$_2$Ca$_{2}$Cu$_3$O$_y$}

\def\8LSCO{La$_{1.88}$Sr$_{.12}$CuO$_4$}
\def\110LNSCO{La$_{1.5}$Nd$_{0.4}$Sr$_{0.1}$CuO$_{4}$}
\def\stage4LCO{La$_{2}$CuO$_{4+\delta}$}
\def\Y248{YBa$_2$Cu$_4$O$_8$}

\def\NbSe2{NbSe$_2$}
\def\TaSe2{TaSe$_2$}
\def\TiSe2{TiSe$_2$}

\begin{document}

\title{Properties of the zero energy Andreev bound state in a two sublattice SNS junction }
 \author{Chandan Setty}
\affiliation{Department of Physics and Institute for Condensed Matter Theory, University of Illinois 1110 W. Green Street, Urbana, IL 61801, USA}
\author{Jiangping Hu}
\affiliation{Beijing National
Laboratory for Condensed Matter Physics, Institute of Physics,
Chinese Academy of Sciences, Beijing 100080,
China}
\affiliation{Department of Physics and Astronomy, Purdue University, West
Lafayette, Indiana 47907, USA}
\begin{abstract}
We study properties of the zero energy Andreev bound state in a superconductor-normal metal-superconductor(SNS) junction consisting of two intrinsic degrees of freedom. The superconductors on either sides of the normal metal are assumed to have two sublattices with an $intra$-sublattice pairing with a phase of zero or $\pi$ between the two sublattices. In addition, we add a uniform $inter$-sublattice pairing and study its effect on the local density of states (LDOS). In particular, we find that as the inter-sublattice pairing is turned on, the zero bias peak (ZBP) is unstable (robust) when the phase difference across the sublattices is $\pi$ (zero). We discuss the relevance of our results to the recently proposed odd parity $-\eta$ pairing ground states in Iron based superconductors (FeSCs).
\end{abstract}

\maketitle

\begin{center}
{\textbf{I.   INTRODUCTION}}
\end{center}
The study of electronic transport through the normal metal-superconductor interface has provided a wealth of information regarding the ground states of the constituent materials forming such an interface. A case of particular interest is the transport of an electron having an energy less than that of the superconducting gap, which, when  incident onto the interface from the normal metal region, gets retro-reflected as a hole with an opposite spin. Such a reflection conserves momentum (upto order $\Delta/E_f$ where $\Delta$ is the gap and $E_f$ is the Fermi energy) and is accompanied by the transmission of a charge $2e$ Cooper pair into the superconductor. This process, termed as Andreev reflection\cite{Andreev1964, BTK1982}, has had a widespread applicability in probing a variety of condensed matter systems for several decades, including superconducting ground states\cite{DeutscherRMP}, polarization in ferromagnets\cite{Beenakker1995, Mazin1999}, and more recently, topological matter \cite{Kane2008,Ng2009,LSS2010}.\\
\newline
Probing the superconducting gap and its pairing symmetry is, perhaps, one of the most important use of Andreev reflection spectroscopy\cite{DeutscherRMP}. The sensitivity of the Andreev in-gap bound state energies to the phase of the superconducting gap gives it a special advantage over probes like ARPES, STM or Raman which do not contain any phase information. As an example, the surface of a $d-$ wave superconductor, when oriented perpendicular to the nodal direction of the order parameter, was predicted to host a zero energy surface bound state\cite{CRHu1994}. Kashiwaya and coworkers\cite{Kashiwaya-Review, Kashiwaya1995, Kajimura1995} generalized the works of references\cite{BTK1982,CRHu1994} to a $d-$ wave symmetry of the gap for arbitrary orientations of the surface. Such ZBPs have since been widely observed in experiments (see refs.\cite{Strasik1998,Ng1998,Revcolevschi2000} and refs. \cite{Kashiwaya-Review,DeutscherRMP} for a review) providing a crucial confirmation of the $d-$ wave pairing gap in the Cuprates. Additionally, the effects of a magnetic field and surface roughness\cite{CRHu1994-Robust,Sauls1997,Stefanakis2002} on the ZBP, and a broken time reversal component\cite{CRHu1994-Robust,Stefanakis2002} of the $d-$ wave order parameter have all been taken into account. Similarly, characteristic features of the Andreev tunneling spectrum have been calculated\cite{Zwicknagl1981} and observed\cite{Lichtenberg2000} in the spin triplet $p-$ wave superconductor $Sr_2 RuO_4$. More recently, a topological index theorem for the stability of dispersionless surface Andreev bound states on the interface of a normal metal-superconductor was proven\cite{Yokoyama2011}. Here, the authors derived general conditions for the existence of dispersionless Andreev bound states for different angular momenta as well as mixed pairing symmetries using the bulk-boundary correspondence. \\
\newline
In multiorbital systems, the resolution  of the pairing symmetry using Andreev and tunneling spectroscopy\cite{Tanaka2013,Kashiwaya2014} is complicated by the fact that several bands contribute to the density of states at the Fermi level. For example, in the  iron based superconductors \cite{Dolgov2009,Sudbo2009,Sudbo2009Andreev,Bernevig2008,Barash2005,Greene2015,Devyatov2014,Mazin2013}, all the iron $d-$ orbitals are known to have a non-vanishing density of states at $E_f$ resulting in multiple pockets around the $\Gamma$ and $M$ points in the Brillouin zone\cite{Scalapino2009}. To add to it, unlike the Cuprates, the phase distribution of the order parameter across the Fermi surface is not symmetry protected; as a result, phase sensitive probes like those used in the Cuprates are hard to design. So far, several new pairing phases in the iron based superconductors have been proposed, including, $s-$ wave\cite{Mazin2008,Kuroki2008,Chubukov2008,Hu2008,Hu2011,Onari2010} and $d-$ wave pairing symmetries\cite{Bernevig2011-dwave, Scalapino2011}. Within the s-wave pairing symmetry, there are a variety of possibilities including the sign changes between different pockets (so called $s^\pm$\cite{Mazin2008,Kuroki2008,Chubukov2008,Hu2008}), or between bands featured by different orbitals\cite{Hu2012-Xiaoli,Hu2014,Hu2013-OddParity,Kotliar2014}, or between two sublattices\cite{Hu2013-OddParity} (the $\eta$ pairing) . Therefore,  a systematic study of these ground state characteristics using Andreev spectroscopy is called for.\\
\newline
\begin{figure}[h!]
\includegraphics[height=0.08\textheight,width=0.45\textwidth]{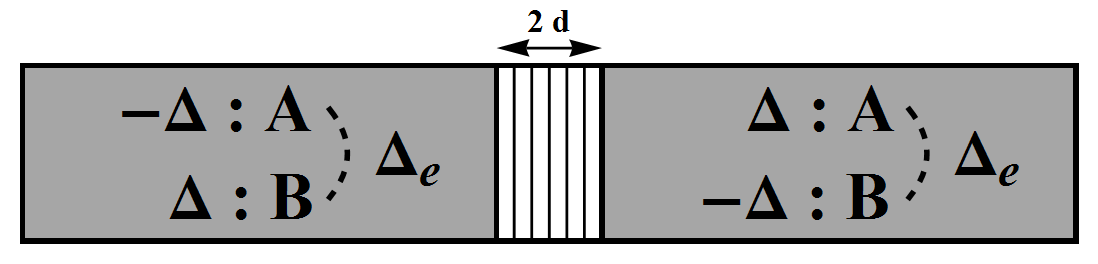}
\caption{\label{SNS2}Figure describing the SNS geometry for the two sub-lattice case. We consider a sign change between the left and right superconductors on each sublattice ($A, B$) and also a sign change between sublattices $A$ and $B$ within a superconductor. Additionally, we have an uniform even parity pairing between $A$ and $B$. A normal metal of thickness $2d$ is sandwiched between the two SCs. The origin is taken to be at the right $NS$ interface}
\end{figure}
The main focus of this work will be to analyze the properties of SNS junctions formed by such exotic superconducting states proposed for multiorbital systems \cite{Hu2014,Hu2013-OddParity}. Before we outline the details of our calculations, we briefly summarize previous work by one of us \cite{Hu2014,Hu2013-OddParity} so that the distinction between the mixed $\eta$ and the commonly studied $s_{\pm}$ pairing states is clarified. In order to accommodate the two crucial experimental features of (i) a $(\pi/2, \pi)$ resonance in the 245 Chalcogenide and (ii) an isotropic, node-less gap on both the electron pockets in most iron superconductors, one needs to include both nearest neighbor (NN) even parity pairing, $\langle c_{\vec k \uparrow}^{\dagger} c_{-\vec k \downarrow}^{\dagger}\rangle$, and the next nearest neighbor (NNN) odd parity $\eta$ pairing, $\langle c_{\vec k \uparrow}^{\dagger} c_{-\vec k+ \vec Q \downarrow}^{\dagger}\rangle$. Here $c_{\vec k\sigma}^{\dagger}$ is the electron creation operator at quasi-momentum $\vec k$ and spin $\sigma$, and $\vec Q=(\pi,\pi)$ is the reciprocal lattice wave vector in the 2-Fe unit cell. While the former brings about the sign change between the inner and outer electron and hole pockets in the two Fe BZ (thus providing a possible explanation of the resonance peak), the latter removes the nodes at the intersection point of the two electron pockets giving a completely node-less gap on the electron pockets. A natural consequence of the odd parity pairing is that if we convert the $\langle c_{\vec k \uparrow}^{\dagger} c_{-\vec k+ \vec Q \downarrow}^{\dagger}\rangle$ order parameter in terms of the pairing in the A and B sublattices (by re-writing $c_{\vec k}$ and $c_{-\vec k+ \vec Q}$ in terms of $c_{A \vec k}$ and $c_{B \vec k}$), we find that the order parameter between the two sub-lattices A and B should change sign. This real space sign distribution of the gap function as well as its momentum space counter part are, therefore, clearly distinct from the $s_{\pm}$ scenario. \\
\newline
To begin our detailed analysis of the bound state properties of an SNS junction formed by a pairing state described above, we model the lattice with two intrinsic degrees of freedom, say two sublattices.  We assume that the superconductors on either side of the normal metal have two kinds of local pairings - type $(a)$ a $NNN$ intra-sublattice pairing with a relative phase of $0$ or $\pi$ between the two sublattices and type $(b)$ a $NN$ uniform inter-sublattice pairing in each of the two superconductors (see figs \ref{SNS2} and \ref{SNS3}). We also maintain an overall $\pi$ phase difference between the two superconductors on either side of the normal metal $-$ a condition which gives a ZBP at the NS interface in the absence of the $(b)$ pairing type. Our central result is that, as the inter-sublattice pairing type $(b)$ is turned on, the ZBP is unstable (robust) when the phase difference between the sublattices is $\pi$ (zero). This result, as will be discussed later, rules out a pure finite momentum $\eta$ pairing. In the following section II, we will present an analytic result for a simplified toy model of an SNS junction. In section III we will discuss our numerical results supporting the conclusions in section II. Finally, in section IV we will end with our discussions and provide experimental context to our results through FeSCs.\\
\begin{center}
{\textbf{II. ANALYTICAL TREATMENT}}
\end{center}
(i)\textit{Sign change case:} We begin with the case where we have two sublattices in each superconductor, labelled by $A$ and $B$. For the superconductor on the right, we assume a positive sign of the gap on $A$ and negative on $B$. On the left, we swap the signs of the gaps leaving the magnitudes the same(see fig \ref{SNS2}). The B-dG equations in the bulk of the right superconductor take the form
\begin{eqnarray}
\epsilon u_1^R &=&  -H_e u_1^R + \Delta v_1^R + \Delta_e v_2^R\\ \nonumber
\epsilon v_1^R &=&  H_e v_1^R + \Delta u_1^R + \Delta_e u_2^R\\ \nonumber
\epsilon u_2^R &=&  -H_e u_2^R - \Delta v_2^R + \Delta_e v_1^R\\ \nonumber
\epsilon v_2^R &=&  H_e v_2^R - \Delta u_2^R + \Delta_e u_1^R.
\end{eqnarray}
Here $u_i^{L,R}, v_i^{L,R}$ are the B-dG wavefunctions for the left (L) and right (R) superconductors, $\Delta,\Delta_e$ are the intra- and inter- sublattice pairing respectively, $H_e$ is the kinetic part of the Hamiltonian and $\epsilon$ is the eigenvalue. We can solve the ensuing differential equations to get the B-dG solutions $u_i^R, v_i^R$ in the right superconductor as
\begin{eqnarray}
u_1^R(x) &=& \alpha_{11}^R e^{i k_1 x} + \alpha_{12}^R  e^{-i k_2 x}\\ \nonumber
u_2^R(x) &=& \alpha_{21}^R e^{i k_1 x} + \alpha_{22}^R  e^{-i k_2 x}\\ \nonumber
v_1^R(x) &=& \beta_{11}^R e^{i k_1 x} + \beta_{12}^R  e^{-i k_2 x}\\ \nonumber
v_2^R(x) &=& \beta_{21}^R e^{i k_1 x} + \beta_{22}^R  e^{-i k_2 x},
\end{eqnarray}
where $\alpha^R s$ and $\beta^R s$ are coefficients on the right superconductor whose relations will be determined below and  $k_1$ and $k_2$ are defined as $k_1 = \sqrt{E_f + \sqrt{\epsilon^2 - (\Delta^2 + \Delta_e^2)}}$ and $k_2 = \sqrt{E_f - \sqrt{\epsilon^2 - (\Delta^2 + \Delta_e^2)}}$.
 The above solutions are substituted into the B-dG equations to obtain a relations between the coefficients $\beta^R s$ and $\alpha^R s$. The matrix equation which relates the two sets is given by
\small
\begin{widetext}
\begin{align}
\begin{bmatrix}
-\Delta e^{i k_1 x} \hfill & -\Delta e^{-i k_2 x} \hfill & -\Delta_e  e^{i k_1 x} \hfill & - \Delta_e  e^{-i k_2 x} \hfill \\
(\epsilon + A) e^{i k_1 x}\hfill & (\epsilon + B) e^{-i k_2 x} \hfill & 0 \hfill & 0 \hfill \\
 -\Delta_e e^{i k_1 x} \hfill & -\Delta_e e^{-i k_2 x} \hfill & \Delta  e^{i k_1 x} \hfill & \Delta  e^{-i k_2 x} \hfill \\
0\hfill & 0 \hfill &(\epsilon + A) e^{i k_1 x}\hfill & (\epsilon + B) e^{-i k_2 x} \hfill l \\
\end{bmatrix}  
\begin{bmatrix}
\beta_{11}^R \hfill  \\
\beta_{12}^R \hfill \\
\beta_{21}^R \hfill\\
\beta_{22}^R \hfill\\
\end{bmatrix}\nonumber 
=
\begin{bmatrix}
-(\epsilon-A) e^{i k_1 x} \alpha_{11}^R - (\epsilon - B) e^{-i k_2 x} \alpha_{12}^R \hfill  \\
\Delta e^{i k_1 x} \alpha_{11}^R + \Delta e^{-i k_2 x} \alpha_{12}^R + \Delta_e e^{i k_1 x} \alpha_{21}^R + \Delta_e e^{-i k_2 x} \alpha_{22}^R\hfill \\ 
-(\epsilon-A) e^{i k_1 x} \alpha_{21}^R - (\epsilon - B) e^{-i k_2 x} \alpha_{22}^R \hfill  \\
-\Delta e^{i k_1 x} \alpha_{21}^R - \Delta e^{-i k_2 x} \alpha_{22}^R + \Delta_e e^{i k_1 x} \alpha_{11}^R + \Delta_e e^{-i k_2 x} \alpha_{12}^R\hfill 
\end{bmatrix},
\end{align}\\
\end{widetext}
\normalsize
where we have defined $A= k_1^2- E_f$ and $B = k _2^2- E_f$. The above sets of equations can be solved and $\beta^R$s are written in terms of $\alpha^Rs$ as
\begin{eqnarray}
\beta_{11}^R &=& e^{-i \phi_2}( \alpha_{11}^R  sin\Xi+  \alpha_{21}^R cos \Xi)\\\nonumber
\beta_{12}^R &=& e^{i \phi_2}( \alpha_{12}^R sin\Xi + \alpha_{22}^R cos \Xi )\\\nonumber
\beta_{21}^R &=& e^{-i \phi_2}( \alpha_{11}^R cos\Xi - \alpha_{21}^R sin \Xi )\\\nonumber
\beta_{22}^R &=& e^{i \phi_2}( \alpha_{12}^R cos\Xi - \alpha_{22}^R sin \Xi ),
\end{eqnarray}
with
\begin{eqnarray}
cos \phi_2 &=& \frac{\epsilon}{\sqrt{\Delta^2 + \Delta_e^2}}\\
cos \Xi &=& \frac{\Delta_e}{\sqrt{\Delta^2 + \Delta_e^2}}. 
\end{eqnarray}
We can follow the same procedure for the case of the left superconductor noting that the sign of the order parameters is swapped. The general solutions of the B-dG equation in the normal metal are given by $u_j^N(x) = p_{j1} sin[k_1' (x+d)] + p_{j2} cos[k_1'(x+d)]$,
where $p_{ij}$ are numbers, $j=1,2$, and similar equations follow for $v_j^N(x)$ with coefficients $q_{ij}$ and $k_1'$ replaced by $k_2'$. We have defined $k_1' = \sqrt{E_f + \epsilon}$ and $k_2' = \sqrt{E_f - \epsilon}$. As a next step, we match boundary conditions on the two sides of the interface at $x=0$ using (the origin is chosen as the interface of the right SN junction, see Fig \ref{SNS2})
\begin{eqnarray}
u_i^N(x=0) &=& u_i^R(x=0)\\  \nonumber
u_{i x}^N(x=0) &=& u_{i x}^R(x=0),\\ \nonumber
\end{eqnarray}

and similar equations follow for $x=2d$ and for the $v_i's$. Here $i=1,2$, and the subscript $x$ refers to derivatives. With these boundary conditions we obtain sixteen equations with sixteen variables to solve for. We can determine the condition for the existence of a solution by equating the determinant of the linear equation matrix to zero, which then yields the condition
\begin{eqnarray}
cos(2 \Xi) - cos[2(k_1' - k_2' -\phi_2)]&=&0,\\ \nonumber
\end{eqnarray}
which implies that
\begin{eqnarray}
 k_1' - k_2' -\phi_2 &=& \Xi + n \pi,\label{SC-Condition}
\end{eqnarray}
where,
\begin{equation}
\Xi = \frac{1}{2} cos^{-1}\left[ \frac{\Delta_e^2 - \Delta^2}{\Delta_e^2 + \Delta^2}\right].
\end{equation}
The ratio of the in gap bound state energy($\epsilon$) and the Fermi energy ($ E_F$) is small, and as a result, the difference of the wavevectors in the normal metal ($k_1' - k_2'$) is proportional to $\epsilon$, which is in turn proportional to $cos \phi_2$. Thus the condition above in eq.\ref{SC-Condition} states that when there is a sign change of $\pi$ between the sublattices, even an $\textit{infinitesimally}$ small inter-sublattice uniform pairing will destroy the ZBP. It is easy to show, by performing a unitary transformation to the Hamiltonian which diagonalizes the pairing sector, that turning on the inter-sublattice pairing is equivalent to the effects of a non-zero inter-sublattice hopping matrix element. 
In the limit of $\Delta_e\ll\Delta$, we keep only the linear terms to get
\begin{equation}
k_1' - k_2' - \phi_2 = \frac{\pi}{2} - \eta + n \pi
\end{equation}
with $\eta \equiv \frac{\Delta_e}{\Delta}$. Thus, the presence of even a small but non-zero $\eta$ leaves the ZBP unstable.
\begin{figure}[h!]
\includegraphics[height=0.08\textheight,width=0.45\textwidth]{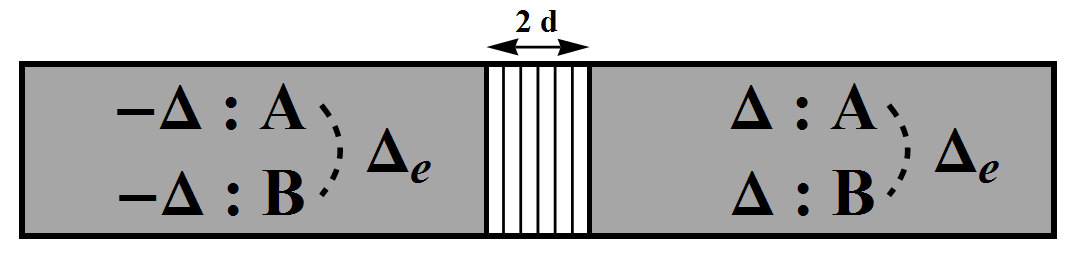}
\caption{\label{SNS3}Figure describing the SNS geometry for the two sub-lattice case with sign change between left and right superconductor, like fig \ref{SNS2}. However, in this case, there is no internal sign change between the two sublattices.}
\end{figure}\\
\newline
(ii)\textit{No sign change case:} We now want to compare our previous result in (i) to the case where there is no internal sign change between sublattices $A$ and $B$ as shown in Fig \ref{SNS3}. Just like before, we wish to study the effect of the nearest neighbor uniform pairing ($\Delta_e$). To this end, the Bd-G equations for the right superconductor of the new geometry are modified as
\begin{eqnarray}
\epsilon u_1^R &=&  -H_e u_1^R + \Delta v_1^R + \Delta_e v_2^R\\ \nonumber
\epsilon v_1^R &=&  H_e v_1^R + \Delta u_1^R + \Delta_e u_2^R\\ \nonumber
\epsilon u_2^R &=&  -H_e u_2^R + \Delta v_2^R + \Delta_e v_1^R\\ \nonumber
\epsilon v_2^R &=&  H_e v_2^R +\Delta u_2^R + \Delta_e u_1^R.
\end{eqnarray}
We can follow the same procedure to obtain the matrix relating the $\alpha^R$s and $\beta^R$s which reads
\small
\begin{widetext}
\begin{align}
\begin{bmatrix}
-\Delta e^{i k_1 x} \hfill & -\Delta e^{-i k_2 x} \hfill & -\Delta_e  e^{i k_1 x} \hfill & - \Delta_e  e^{-i k_2 x} \hfill \\
(\epsilon + A) e^{i k_1 x}\hfill & (\epsilon + B) e^{-i k_2 x} \hfill & 0 \hfill & 0 \hfill \\
 -\Delta_e e^{i k_1 x} \hfill & -\Delta_e e^{-i k_2 x} \hfill & -\Delta  e^{i k_1 x} \hfill & -\Delta  e^{-i k_2 x} \hfill \\
0\hfill & 0 \hfill &(\epsilon + A) e^{i k_1 x}\hfill & (\epsilon + B) e^{-i k_2 x} \hfill l \\
\end{bmatrix}  
\begin{bmatrix}
\beta_{11}^R \hfill  \\
\beta_{12}^R \hfill \\
\beta_{21}^R \hfill\\
\beta_{22}^R \hfill\\
\end{bmatrix}\nonumber 
=
\begin{bmatrix}
-(\epsilon-A) e^{i k_1 x} \alpha_{11}^R - (\epsilon - B) e^{-i k_2 x} \alpha_{12}^R \hfill  \\
\Delta e^{i k_1 x} \alpha_{11}^R + \Delta e^{-i k_2 x} \alpha_{12}^R + \Delta_e e^{i k_1 x} \alpha_{21}^R + \Delta_e e^{-i k_2 x} \alpha_{22}^R\hfill \\ 
-(\epsilon-A) e^{i k_1 x} \alpha_{21}^R - (\epsilon - B) e^{-i k_2 x} \alpha_{22}^R \hfill  \\
\Delta e^{i k_1 x} \alpha_{21}^R + \Delta e^{-i k_2 x} \alpha_{22}^R + \Delta_e e^{i k_1 x} \alpha_{11}^R + \Delta_e e^{-i k_2 x} \alpha_{12}^R\hfill 
\end{bmatrix},
\end{align}\\
\end{widetext}
\normalsize
where the new definitions of the wave vectors in the superconductor on the right are given by $k_1 = \sqrt{E_f + \sqrt{\epsilon^2 - (\Delta \pm \Delta_e)^2}}$ and $k_2 = \sqrt{E_f - \sqrt{\epsilon^2 - (\Delta \pm \Delta_e)^2}}$,
and the remaining definitions remain unchanged. We again solve the simultaneous equations for the coefficients $\beta^R$s in terms of $\alpha^R$s on the right superconductor to get the new equations in terms of hyperbolic functions instead of harmonic functions -
\begin{eqnarray}
\beta_{11}^R &=& -\gamma_1 (\alpha_{11}^R cosh \Gamma  - \alpha_{21}^R sinh \Gamma)\\ \nonumber
\beta_{12}^R &=& -\gamma_2 (\alpha_{12}^R cosh \Gamma  - \alpha_{22}^R sinh \Gamma)\\ \nonumber
\beta_{21}^R &=& \gamma_1 (\alpha_{11}^R sinh \Gamma  - \alpha_{21}^R cosh \Gamma)\\ \nonumber
\beta_{22}^R &=& \gamma_2 (\alpha_{12}^R sinh \Gamma  - \alpha_{22}^R cosh \Gamma)\\ \nonumber
\end{eqnarray}
with the definitions
\begin{eqnarray}
\gamma_1 &=& \frac{A- \epsilon}{\sqrt{\Delta^2 - \Delta_e^2}} \equiv i \alpha + \beta, \\ 
\gamma_2 &=& \frac{B- \epsilon}{\sqrt{\Delta^2 - \Delta_e^2}} \equiv -i \alpha + \beta, \\ 
cosh \Gamma &=& \frac{\Delta}{\sqrt{\Delta^2 - \Delta_e^2}}, 
\end{eqnarray}
where $\alpha$ and $\beta$ are real in the case $\Delta_e<\Delta$. In the limit $\eta = \frac{\Delta_e}{\Delta}\ll1$ we have 
\begin{eqnarray}
\alpha &\approx& sin \phi + \frac{\eta}{sin \phi},\\
\beta &\approx& - cos \phi,\\
1 &\approx& cosh 2 \Gamma,
\end{eqnarray}
with $\phi$ defined as $cos \phi = \epsilon/ \Delta$. For the case of the left superconductor, the relations between the $\beta^L$ s and $\alpha^L$ s get modified by the substitutions $\gamma_1 \leftrightarrow \gamma_2$ and $ cosh \Gamma \rightarrow - cosh \Gamma$. We can then match boundary conditions at the two interfaces at $x=0$ and $x= -2d$ and we end up with sixteen equations and sixteen variables. The existence of a non-trivial solution requires that the determinant of the homogenous matrix equation is zero and this yields the condition
\begin{eqnarray}
0&=&(\beta^2 - \alpha^2) cos[2(k_1' - k_2')] + (\alpha^2 + \beta^2) cosh 2\Gamma \\ \nonumber
& &- 2 \alpha \beta sin[2(k_1' - k_2')].
\end{eqnarray}
In the limit $\eta\ll1$ we can rewrite the condition as
\begin{eqnarray}
0&=&2 cos^2[k_1' -k_2' - \phi] - 2 \eta cos[2(k_1' - k_2')] + 2 \eta \\ \nonumber
& &+ 2 \eta cot \phi sin[2(k_1' -k_2')]
\end{eqnarray}
where, again, $k_1' - k_2'$ is proportional to $\Delta cos\phi$. The above condition allows for a zero energy bound state solution (i.e in the limit of $\phi \rightarrow \pi/2$), implying that the ZBP is $\textit{robust}$ under a small perturbation $\Delta_e$. However, as one must expect, the ZBP is destroyed in the other limit where $\Delta_e\gg \Delta$.\\
\begin{figure}[h!]
\includegraphics[height=0.18\textheight,width=0.51\textwidth]{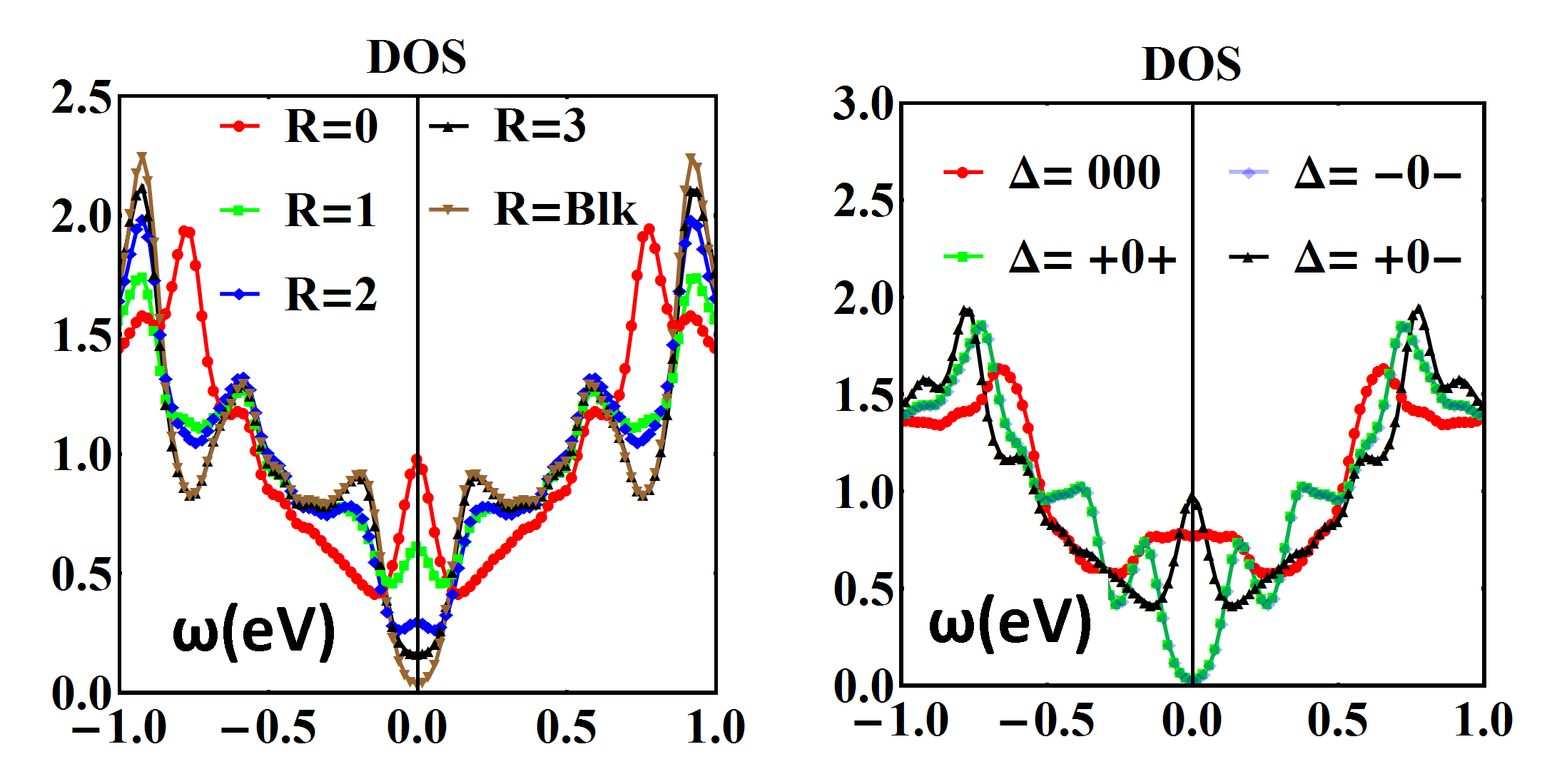}
\caption{ \label{Check}LDOS as a function of the bias voltage. Left: As a function of the position of the tip at different points along a direction perpendicular to the SNS boundary (see also  Fig 3 for the lattice numbering) for the $+NN-$ geometry. Right: As a function of different signs of gap across the interface. }
\end{figure}
\begin{center}
{\textbf{III. NUMERICS}}
\end{center}
a)\textit{ Model and method:} We now verify the conclusions in the previous section using a more concrete, two dimensional, lattice model. For the purposes of numerical illustration, we start with a two sublattice model consisting of a single orbital per sublattice. A model ideally suited for such a calculation is a single subspace of the doublet model presented in \cite{Hu2012-S4} for FeSCs. The parameters are chosen in such a way that there are only hole pockets centered around $\Gamma$ point. We do this to be able to obtain a fully gapped spectrum in the bulk of the superconductor; for the chosen model, this is not possible with a pure type $(a)$ pairing (with a $\pi$ phase between $A$ and $B$) when the electron pockets close to the edges of the Brillouin zone are present. If we insist that such a band structure and Fermi surface must indeed describe a specific system, we can keep in mind the case of the extremely hole doped pnictide $KFe_2As_2$, where ARPES sees only hole pockets at the $\Gamma$ point \cite{Ding2009-HolePockets}. However, as we saw in the earlier section, the results are more generic and can be applied to a wide category of systems. The following sets of parameters are chosen in the two sublattice model\cite{Hu2012-S4}: $(t_{1s}, t_{1d}, t_{2s}, t_{2d}, t_{3s}, t_{3d},\mu, t_c) = (0.2, -0.03, 0.3, 0.2, 0.05, -0.05, 1.8, 0)$ with all units specified in $eV$. The simple Fermi surface that is assumed from this parameterization, allows for a proper gap structure (without nodal behavior) in the bulk even when there is no type $(b)$ pairing. As a result, a numerical study of the in gap bound states and the effect of type $(b)$ pairing on them can be perfomed without ambiguities.\\
\newline
\begin{figure}[h!]
\includegraphics[height=0.18\textheight,width=0.51\textwidth]{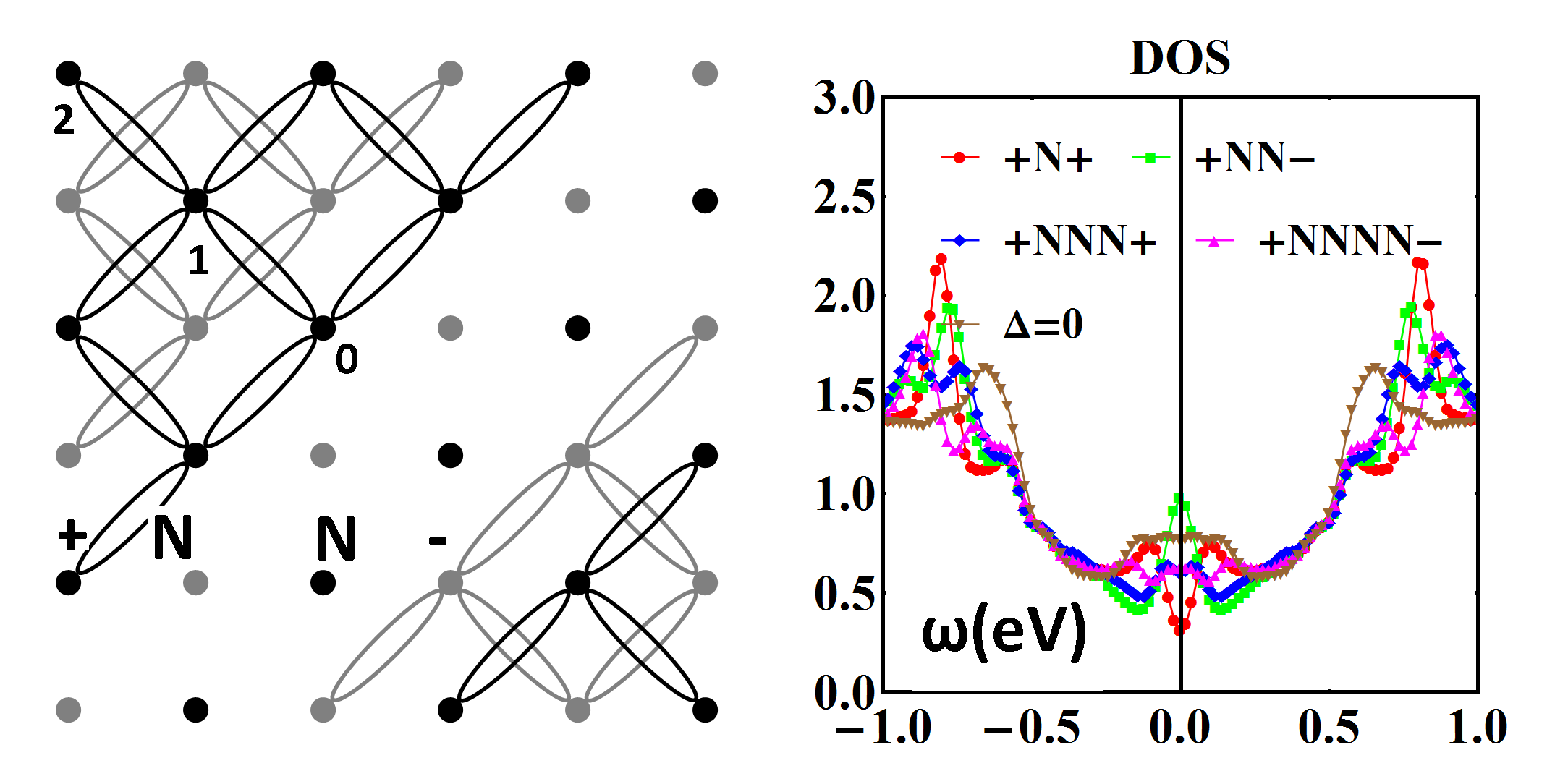}
\caption{\label{SNS}Left: The diagonal SNS junction with the sign changing pairing pattern for the $+NN-$ geometry. The black(gray) denote two different sublattices with positive(negative) pairing. Right: Comparison of the LDOS for the different geometries depending on the number of layers in the normal state.  }
\end{figure}
To calculate the Andreev-LDOS we start with the B-dG equations in real space given by
\begin{eqnarray}
 u_n(r) \epsilon_n &=&\int d \vec{r}'[ H_e(r, r') u_n(r')  + \Delta(r, r') v_n(r')]  \\ \nonumber
 v_n(r) \epsilon_n &=& \int d \vec{ r}'[ -H_e(r, r')^*  v_n(r')  + \Delta(r, r')^* u_n(r')] 
\end{eqnarray}
where $H_e(r, r')$ contains the kinematics in real space in the form of single orbital $NN$ , $NNN$ and 3rd $NN$ hopping. The gap structure in the superconducting state, $\Delta(r,r')$, corresponds to $NNN$ intra-sublattice pairing which changes sign between the sublattices (pure type $(a)$ pairing) with no $NN$ pairing initially. The matrix in real space is diagonalized on a 40 $\times$ 40 lattice and the B-dG factors, $v_n(r)$ and $u_n(r)$, are obtained along with the eigenvalues $\epsilon_n$. One can then go on to obtian the LDOS at $r$ as
\begin{eqnarray}
\rho_r(E) &=& -2 \sum_n [ \mid u_n (r)\mid^2 f'( E - \epsilon_n)  \\ \nonumber
&& +  \mid v_n(r)\mid^2 f'(E + \epsilon_n) ]
\end{eqnarray}
where $f$ is the fermi function
and the prime denotes the derivative w.r.t energy. We fix our temperture to $k_BT = 0.03 eV$  throughout the work. For the two sublattice, single orbital model band structure parameters with a purely type $(a)$ pairing (with $\pi$ phase difference), a relatively large value of the intra-sublattice pairing parameter ($\Delta_0 = 0.4-0.5$) is needed to open a considerable full gap ($\sim 0.08 eV$) on the fermi surface. In the subsection to follow, we will study the properties of such a model without type $(b)$ pairing to verify known results. Later, we will turn on type $(b)$ pairing and study its effects on the ZBP.\\
\newline
b) \textit{Results:} As a warm up, we begin by reproducing already known results that would be expected from a pure type $(a)$ pairing (with a phase difference of $\pi$ between $A$ and $B$). As a start, we calculate the LDOS as a function of the position near the interface. An illustration of the junction is shown Fig \ref{SNS} . A section of the $Fe$ layer is made non-superconducting in the diagonal (110) direction by depositing two atomic layers of a normal metal. This yields opposite signs of the superconducting order parameter on either side of the normal metal. 
The LDOS in Fig \ref{Check} (Left) is calculated along a direction perpendicular to the interface (as shown in Fig \ref{SNS}, marked by numbers 0,1,2). Close to the interface, a ZBP appears, and is suppressed as we move away from the boundary and into the bulk. The origin of the ZBP is a boundary effect originating from the sign change of $\pi$ in the superconducting order parameter right across the interface. To confirm this numerically, we plot the LDOS (fig \ref{Check} Right) as a function of the relative sign change of the order parameter across the interface but keeping the pairing in the bulk intact. As is seen, there is a ZBP only in the case of a sign change of $\pi$ across the interface. Understanding the origin of such a ZBP, involves a change in the sign of the ratios of the B-dG wavefunction solutions, $u_n(r)/v_n(r)$ (obtained self-consistently), across the two $NS$ boundaries which gives rise to an extra ($\pi/2$) phase in the condition for the existence of a solution. This extra phase of $\pi/2$ is responsible for a zero energy state \cite{DeutscherRMP}. A generalized version of this condition in the presence of type $(b)$ pairing is our main contribution in eq.\ref{SC-Condition}.
\newline
\newline
Fig \ref{SNS} (Right) plots a comparison of the LDOS for different SNS geometries obtained by depositing multiple normal metal layers along the (110) direction. The plot on the left of Fig \ref{SNS} shows the geometry for the case of two normal metal layers yielding a $+NN-$ junction ($N$ stands for normal metal layer and the $\pm$ stands for sign of order parameter). Similarly one can obtain a $+NNN+$ junction by depositing three atomic normal metal layers along the (110) direction, and so on. As is expected from the previous arguments based on the phase difference of $\pi$ across the interface, we see a ZBP only when we have an even number of normal metal layers between the superconductors (see fig \ref{SNS} Right).  In addition, we note that the intensity of the ZBP for the case of four layers is drasitically reduced when compared to the case of two layers; this is because of the dependence of the intensity of the ZBP on the ratio $d/\xi$, where $d$ is the width between superconductors and $\xi$ is the coherence length. For comparison, the spectrum in the uniform non-superconducting state is also plotted.
\newline
\newline
So far, we have only considered the effect of a intra-sublattice ($NNN$) pairing which changes sign between the two sublattices (i.e type $(a)$ pairing with phase of $\pi$). We will now turn on the $NN$ inter-sublattice uniform pairing (type $(b)$) and consider its effect on the LDOS, and particularly, its effect on the ZBP. The B-dG equations are similar to what was  presented before, except for additional terms which connect the particle-hole sections of the Hamiltonian generated from the nearest neighbor pairing (denoted by $\Delta_e$). Fig \ref{EvenParityEffect} shows a plot of the LDOS at the $NS$ boundary of the $+NN-$ geometry. The plot on the left shows the effect of $\eta = \frac{\Delta_e}{\Delta}$ on the ZBP for the case where there is a sign change of $\pi$ between the $A$ and $B$ sublattices. Evidently, upto the accuracy of the given lattice size, even a small value of $\eta = 0.08$ is enough to destroy the ZBP. By choosing larger system sizes, this effect can be seen more clearly with even smaller values of $\eta$. On the other hand, when there is no sign change between the sublattices (Fig\ref{EvenParityEffect} Right), even a relatively large value of $\eta = 0.4$ does not destroy the ZBP. These numerical results, presented for the two dimensional lattice model, are consistent with our analytical result derived in the previous section. In the following, we will discuss the relevance of our results in the context of FeSCs.
\newline
\begin{center}
{\textbf{V.  DISCUSSION}}
\end{center}
Recently, several new pairing phases in the FeSCs have been proposed including $s-$ wave\cite{Mazin2008,Kuroki2008,Chubukov2008,Hu2008,Hu2011,Onari2010} and $d-$ wave pairing symmetries\cite{Bernevig2011-dwave, Scalapino2011}. Additionally, within the s-wave pairing symmetry, there are a variety of possibilities including the sign changes between different pockets (so called $s^\pm$\cite{Mazin2008,Kuroki2008,Chubukov2008,Hu2008}), or between bands featured by different orbitals\cite{Hu2012-Xiaoli,Hu2014,Hu2013-OddParity,Kotliar2014}, or between two sublattices\cite{Hu2013-OddParity}.
  For example, it was argued in \cite{Hu2012-S4} that the $Fe$ lattice in the superconducting state can be divided into two sublattices, with the pairing amplitude changing sign between the two sublattices. The motivation of such state\cite{Hu2013-OddParity, Hu2012-S4} is to gap out the regions where the electron pockets are degenerate so as to be consistent with ARPES.
 However a pure $NNN$ sublattice sign changing gap alone could not account for the experimentally observed neutron resonance peak close to $(\pi, \pi/2)$. An additional $NN$ uniform pairing between the two sublattices was needed for such a sign changing ground state to be fully consistent with experiments. From the above theoretical analysis and the following experimental evidence, we argue that Andreev bound state spectroscopy also requires the presence of $NN$ pairing as was done by neutron scattering, thus ruling out a pure $NNN$ sign changing ground state description of FeSCs. Several experimental groups have reported Andreev spectra in the 1111 \cite{Buchner2009-1111, Canfield2009-1111, Chien2008-Nature, Chien2009-1111, Cohen2009-1111, Gonnelli2010-SST, Gonnelli2011-Review, Wen2008-1111, Wen2009-1111, Karpinski2009-1111, Khlybov2011-1111, Kim2009-1111, Kremer2009-1111, Kulikova2013, Zhao2008-Sm1111, Zhao2008-Nd1111}, 111 \cite{Takeuchi2012-111}, 11 \cite{ Wolf2011-11, Kulikova2013} and 122 \cite{Canfield2009-122, Greene2010-122, Gonnelli2010-122, Gonnelli2010-SST, Gonnelli2011-Review, Samuely2009-122, Takeuchi2010-122} families. Initially, there were several reports of the presence of a ZBP in the 1111 and 122 compounds. However, there were inconsistencies present in the data like dependence of the ZBP on the sample position, temperature dependence, and dependence on the size of the metallic tip. Later it was shown that the ZBP was an artifact of excessive pressure applied on the tip. More consistent data followed once 'soft contact' $Ag$ tips were used (For a detailed review and discussion, the reader can refer to \cite{Gonnelli2011-Review}). In the 1111 compounds ($LaFeAsO_{1-x}F_x$ and $SmFeAsO_{1-x}F_x$) the data was collected for polycrystalline samples with random orientations. No ZBP was observed in either of the cases and as a result it was concluded that no nodes were present and the gaps were completely isotropic \cite{Gonnelli2011-Review} which - if it is to be understood in the odd parity picture - corresponds to the presence of the $NN$ uniform even parity pairing. With respect to the 122 compounds, ($(Ba,K)Fe_2As_2$ and $(Ba,Co)Fe_2As_2$) directional measurements along different crystal directions could be performed due to the avialability of single crystals.
\begin{figure}[h!]
\includegraphics[height=0.18\textheight,width=0.52\textwidth]{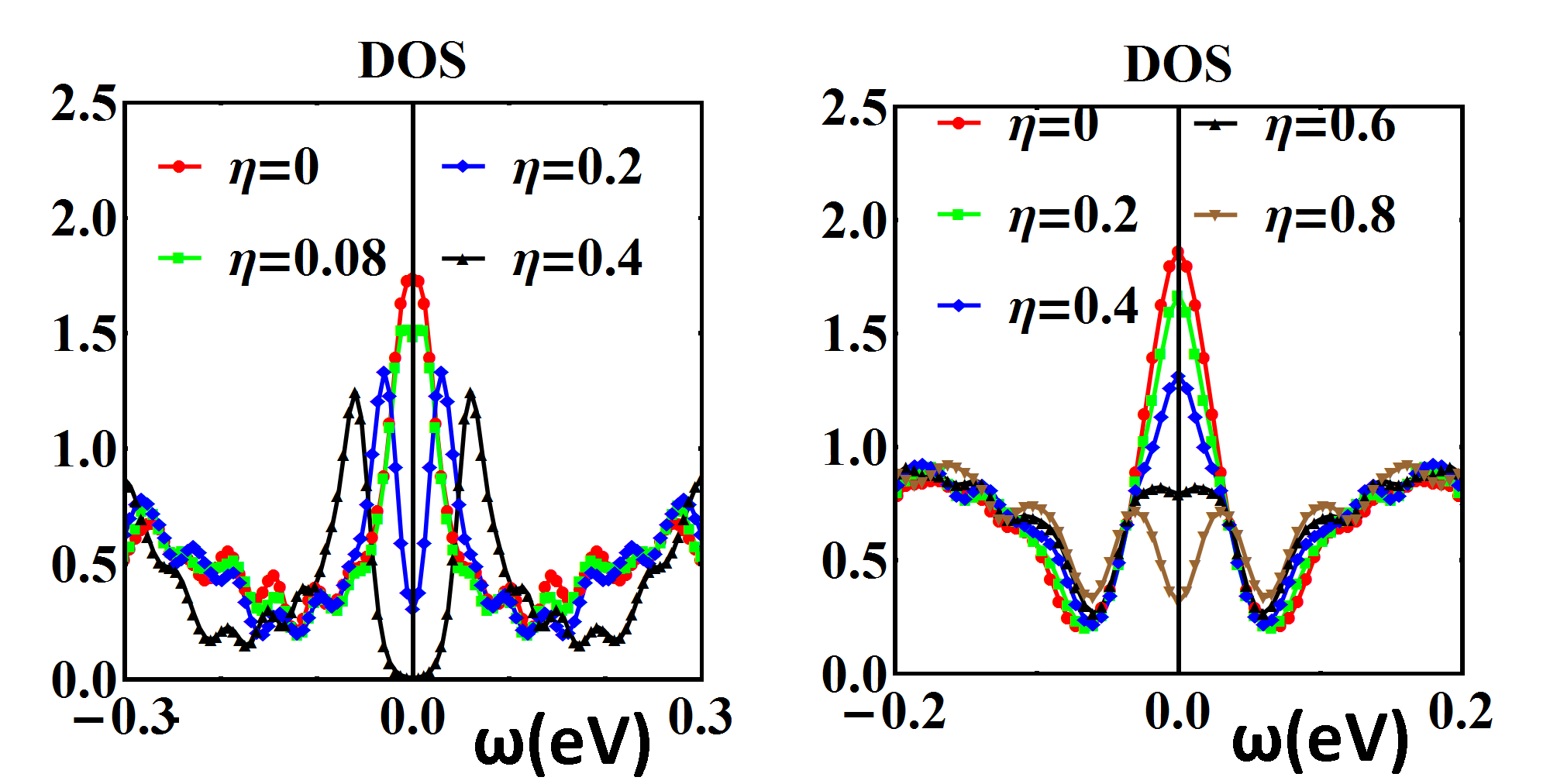}
\caption{\label{EvenParityEffect}Effect of NN even parity pairing on the ZBP in the LDOS. (Left) Case where there is a sign change of $\pi$ between the two sublattices. (Right) Case where there is no sign change between the two sublattices. $\eta\equiv\frac{\Delta_e}{\Delta} $. The ZBP is quickly destroyed in the former case. }
\end{figure}
Several measurements in different directions along the $a-b$ plane again indicated the absence of ZBPs (see \cite{Gonnelli2011-Review} and references therein). Point contact measurements along the $c-$ axis of single crystals in the 111 compound $LiFeAs$ \cite{Takeuchi2012-111} showed a zero bias Josephson current, unrelated to the ZBP due to in gap states. Similar behavior was seen in polycrystalline samples of the 11 compound $FeSe$ \cite{Wolf2011-11, Kulikova2013}. With such strong experimental support and our theoretical analysis,  a pure $NNN$ sublattice sign changing pairing, namely a pure $\eta$-pairing state \cite{Hu2013-OddParity}, is ruled out. But a mixed pairing  state  with both the $\eta$-pairing and a $NN$ uniform pairing is still possible.\\ 
\newline
To conclude, we have studied the properties of the zero bias in gap Andreev bound state for an SNS junction made of a  lattice with two degress of freedom. The superconducting state consists of a $NNN$ pairing with opposite (same) signs on each sublattice and a uniform $NN$ pairing between the two. We find that the zero bias Andreev bound state is unstable (robust) to inter-sublattice $NN$ uniform pairing when the phase difference between the two sublattices is $\pi$ (zero). From existing experimental evidence for FeSCs, this necessarily means that a pure $\eta$ pairing is ruled out. Our study suggests that it is difficult to detect a real space pairing sign change without  symmetry protection in SNS junctions. These results can also be extended to cases for pairing states with sign change among intrinsic degree of freedoms such as orbital. \\
\newline
\textit{Acknowledgements:} CS is supported by the Center for
Emergent Superconductivity, a DOE Energy Frontier Research Center,
Grant No. DE-AC0298CH1088. Hospitality provided by the Institute of Physics, Chinese Academy of Sciences is greatly appreciated.

\bibliographystyle{apsrev}
\bibliography{Andreev}

\end{document}